\documentclass{elsart4-1}

\usepackage{epsfig}
\usepackage{amssymb}

\usepackage[english,francais]{babel}

\newtheorem{e-proposition}[theorem]{Proposition}

\newtheorem{e-definition}[theorem]{Definition\rm}

\renewcommand{\theequation}{\arabic{equation}}
\def\og{\leavevmode\raise.3ex\hbox{$\scriptscriptstyle\langle\!\langle$~}}
\def\fg{\leavevmode\raise.3ex\hbox{~$\!\scriptscriptstyle\,\rangle\!\rangle$}}

\begin{document}

\begin{frontmatter}


\selectlanguage{english}
\title{Newtonian gravity in $d$ dimensions }

\vspace{-4cm}
\selectlanguage{francais}
\title{ }

\selectlanguage{english}
\author[authorlabel1]{Chavanis Pierre-Henri}
\ead{chavanis@irsamc.ups-tlse.fr}

\address[authorlabel1]{Laboratoire de Physique Th\'eorique, Universit\'e Paul Sabatier, 118 route de Narbonne, 31062 Toulouse, France}

\begin{abstract}
We study the influence of the dimension of space on the
thermodynamics of the classical and quantum self-gravitating gas. We
consider Hamiltonian systems of self-gravitating particles described
by the microcanonical ensemble and self-gravitating Brownian
particles described by the canonical ensemble. We present a gallery
of caloric curves in different dimensions of space and discuss the
nature of phase transitions as a function of the dimension $d$. We
also provide the general form of the Virial theorem  in $d$
dimensions and discuss the particularity of the dimension $d=4$ for
Hamiltonian systems and the dimension $d=2$ for Brownian systems.



\end{abstract}
\end{frontmatter}

\selectlanguage{english}
\section{Introduction}
\label{}

Self-gravitating systems have a strange thermodynamics due to the
attractive long-range nature of the gravitational interaction
\cite{paddy,houches,gross,vega}. This gives rise to inequivalence of
statistical ensembles associated with regions of negative specific
heats in the microcanonical ensemble and to phase transitions
associated with gravitational collapse (see, e.g.,
\cite{fermions}). The form of the caloric curve and the nature of the
phase transitions crucially depend on the dimension of space $d$
\cite{sc,fermionsd}. In this paper, we study the influence of the
dimension of space on the thermodynamics of self-gravitating
systems. We distinguish between Hamiltonian and Brownian systems
\cite{hb}. A Hamiltonian system of self-gravitating particles evolves
at fixed mass and energy and is described by the microcanonical
ensemble \cite{paddy}. In the mean-field approximation valid in a
proper thermodynamic limit $N\rightarrow +\infty$ with fixed
$\Lambda=-ER/GM^{2}$ (where $R$ is the size of the system), the
distribution function maximizes, at statistical equilibrium, the
Boltzmann entropy at fixed mass and energy.  The relaxation towards
statistical equilibrium is governed by the mean-field Landau-Poisson
system which conserves mass and energy and monotonically increases the
Boltzmann entropy ($H$-theorem). A system of self-gravitating Brownian
particles
\cite{crs} evolves at fixed mass and temperature and is described by
the canonical ensemble. In the mean-field approximation valid in a
proper thermodynamic limit $N\rightarrow +\infty$ with fixed
$\eta=\beta GMm/R$, the distribution function minimizes, at
statistical equilibrium, the Boltzmann free energy at fixed mass.  The
relaxation towards statistical equilibrium is governed by the
mean-field Kramers-Poisson system which conserves mass and
monotonically decreases the Boltzmann free energy.

These two systems (Hamiltonian and Brownian) are defined in
Sec. \ref{sec_sghb} starting from $N$-body equations. In
Sec. \ref{sec_virial}, we derive the general form of Virial theorem in
$d$ dimensions and we discuss the particularities of the dimension
$d=4$ for Hamiltonian systems and the dimension $d=2$ for Brownian
systems. In the first case, we show that the system evaporates for
$E>0$ and collapses for $E<0$.  In the second case, the system
evaporates for $T>T_{c}$ and collapses for $T<T_{c}$ where
$T_{c}=(N-1)Gm^{2}/4 k_{B}$ is a critical temperature. We also derive
a generalization of the Einstein relation including self-gravity in
$d=2$. Finally, in Sec. \ref{sec_thermo}, we present a gallery of
caloric curves in various dimensions of space for self-gravitating
classical and quantum particles (fermions) and discuss the nature of
phase transitions as a function of $d$.

\section{Self-gravitating Hamiltonian and Brownian systems}
\label{sec_sghb}

In a space of dimension $d$, we consider a system of $N$ particles
with mass $m_{\alpha}$ in gravitational interaction whose dynamics is
described by the equations of motion
\begin{equation}
\ddot x_{i}^{\alpha}=\sum_{\beta\neq \alpha}
{Gm_{\beta}(x_{i}^{\beta}-x_{i}^{\alpha})\over |{\bf
r}_{\beta}-{\bf r}_{\alpha}|^{d}}-\xi_{\alpha} \dot
x_{i}^{\alpha}+\sqrt{2D_{\alpha}}R_{i}^{\alpha}(t). \label{hb1}
\end{equation}
Here, the greak letters refer to the particles and the latin letters
to the coordinates of space.  When the last two terms are set equal
to zero, we recover the usual Hamiltonian model of self-gravitating
systems. However, we consider here a more general situation where
the particles are subject, in addition to self-gravity, to a
friction force and a stochastic force. These terms can mimick the
influence of a thermal bath of non-gravitational origin. We thus
obtain a model of self-gravitating Brownian particles \cite{crs}
extending the usual Brownian model introduced by Einstein and
Smoluchowski. We consider the case of a multicomponent system
\cite{sopik}.  Here, $\xi_{\alpha}$ is the friction coefficient,
$D_{\alpha}$ is the diffusion coefficient and ${\bf R}^{\alpha}(t)$
is a white noise acting independently on the particles and
satisfying $\langle R^{\alpha}_{i}(t)\rangle={0}$ and $\langle
R^{\alpha}_{i}(t)R^{\beta}_{j}(t')\rangle=\delta_{ij}\delta_{\alpha\beta}\delta(t-t')$.
The diffusion coefficient and the friction force satisfy the
Einstein relation $D_{\alpha}=\xi_{\alpha} k_{B}T/m_{\alpha}$ where
$T$ is the temperature of the bath (see below). Since $D\sim T$, the
temperature measures the strength of the stochastic force.

The evolution of the $N$-body distribution function $P_{N}({\bf r}_{1},{\bf v}_{1},...,{\bf r}_{N},{\bf v}_{N},t)$ is governed by the $N$-body Fokker-Planck equation \cite{hb}:
\begin{equation}
\label{hb2} {\partial P_{N}\over\partial t}+\sum_{\alpha=1}^{N}\biggl
({\bf v}_{\alpha}\cdot {\partial P_{N}\over\partial {\bf
r}_{\alpha}}+{{\bf F}_{\alpha}\over m_{\alpha}}\cdot {\partial P_{N}\over\partial {\bf
v}_{\alpha}}\biggr )=\sum_{\alpha=1}^{N} {\partial\over\partial {\bf
v}_{\alpha}}\cdot \biggl\lbrack D_{\alpha}{\partial P_{N}\over\partial
{\bf v}_{\alpha}}+\xi_{\alpha} P_{N}{\bf v}_{\alpha}\biggr\rbrack,
\end{equation}
where ${\bf F}_{\alpha}=-\nabla_{\alpha}U({\bf
r}_{1},...,{\bf r}_{N})$ is the force acting on the
$\alpha$-th particle ($U$ is the potential of
interaction). If we enclose the system within a box and replace the
bare gravitational potential by a soften potential that is regularized
at short distances (so that a statistical equilibrium state exists),
we find that the stationary solution of Eq. (\ref{hb2}) is the canonical
distribution
\begin{equation}
\label{hb3} P_{N}({\bf r}_{1},{\bf v}_{1},...,{\bf r}_{N},{\bf v}_{N})={1\over
Z(\beta)}e^{-\beta H({\bf r}_{1},{\bf v}_{1},...,{\bf r}_{N},{\bf v}_{N})},
\end{equation}
where $H=\sum_{\alpha} m_{\alpha}{v_{\alpha}^{2}\over 2}+U({\bf
r}_{1},...,{\bf r}_{N})$ is the Hamiltonian, provided that the
diffusion and friction coefficients are related to each other by the
Einstein relation. When $\xi=D=0$, Eq. (\ref{hb2}) becomes the
Liouville equation appropriate to Hamiltonian systems. In that case,
Eq. (\ref{hb3}) is not valid anymore. Since energy is now conserved
during the evolution, the system is expected to reach, for
$t\rightarrow +\infty$, the microcanonical distribution
\begin{equation}
\label{hb4} P_{N}({\bf r}_{1},{\bf v}_{1},...,{\bf r}_{N},{\bf v}_{N})={1\over
g(E)} \delta(E-H({\bf r}_{1},{\bf v}_{1},...,{\bf r}_{N},{\bf v}_{N})),
\end{equation}
expressing the equiprobability of accessible microstates (with the
right value of energy).

In the mean-field approximation valid in a proper thermodynamic
limit with $N\rightarrow +\infty$ \cite{hb,sopik}, the evolution of
the distribution function $f_{a}({\bf r},{\bf v},t)$ of each species
of the self-gravitating Brownian gas is governed by the
multicomponent Kramers-Poisson (KP) system
\begin{equation}
\label{hb5}
\frac{\partial f_{a}}{\partial t} + \textbf{v}\cdot\frac{\partial
f_{a}}{\partial \textbf{r}} -\nabla\Phi\cdot\frac{\partial f_{a}}{\partial \textbf{v}} = \frac{\partial}{\partial\textbf{v}}\cdot
\left(D_{a}\frac{\partial f_{a}}{\partial \textbf{v}} + \xi_{a}f_{a}\textbf{v}\right),
\end{equation}
\begin{equation}
\label{hb6}
\Delta\Phi=S_{d}G\int f d{\bf v},
\end{equation}
where $f=\sum_{a}f_{a}$ is the total distribution function. The multicomponent KP system conserves the mass $M_{a}=\int f_{a}d{\bf r}d{\bf v}$ of each species and monotonically decreases the free energy $F=E-TS$ where
\begin{equation}
\label{hb7}
E=\frac{1}{2}\int f v^2\,d{\bf r}\,d{\bf v} +\frac{1}
{2}\int \rho \Phi\,d{\bf r}=K+W,
\end{equation}
is the total energy and
\begin{equation}
\label{hb8} S = - k_B \sum_{a= 1}^{X} \int \frac{f_{a}}{m_{a}}
\ln\left(\frac{f_{a}}{m_{a}}\right) \,d{\bf r}\,d{\bf v},
\end{equation}
is the Boltzmann entropy. The linearly dynamically stable stationary solution of the KP system  is the mean-field Maxwell-Boltzmann distribution
\begin{equation}
\label{hb9}
f_{a}(\textbf{r},\textbf{v}) = A_{a} e^{-\beta m_{a}\lbrack\frac{v^{2}}{2}
+\Phi(\textbf{r})\rbrack},
\end{equation}
that is a local minimum of free energy at fixed mass (canonical
description). In the strong friction limit $\xi\rightarrow +\infty$,
the velocity distribution of the particles is close to the
Maxwellian and the evolution of the spatial distribution is
governed by the multicomponent Smoluchowski-Poisson (SP) system
\cite{hb,sopik}. Alternatively, for $D=\xi=0$ (Hamiltonian systems),
the evolution of the system for $N\rightarrow +\infty$ is governed
by the Vlasov equation \cite{hb}:
\begin{equation}
\label{hb10}
\frac{\partial f}{\partial t} + \textbf{v}\cdot\frac{\partial
f}{\partial \textbf{r}} -\nabla\Phi\cdot\frac{\partial f}{\partial \textbf{v}} = 0.
\end{equation}
The Vlasov equation conserves mass and energy but also the entropy
(and more generally an infinite class of functionals called the
Casimirs). Furthermore, it admits an infinite number of stationary
states (not necessarily Boltzmannian) that can be selected by the
dynamics as a result of an incomplete collisionless violent
relaxation on the coarse-grained scale \cite{lb,next05}. The
statistical equilibrium state in the microcanonical ensemble is
selected by finite $N$ effects (granularities) accounting for
correlations between particles due to close encounters. The
collisional relaxation of stellar systems is usually described by
the mean-field Landau-Poisson system \cite{hb}. In $d=3$ the
multicomponent Landau equation can be written
\begin{eqnarray}
\label{hb11}
\frac{\partial f_{a}}{\partial t} +
\textbf{v}\cdot\frac{\partial f_{a}}{\partial \textbf{r}} -\nabla\Phi\cdot\frac{\partial f_{a}}{\partial \textbf{v}}
= \frac{\partial}{\partial v^{\mu}} \sum_{b = 1}^{X} \int
K^{\mu \nu} \left(m_{b} f'_{b}
\frac{\partial f_{a}}{\partial v^{
\nu}} - m_{a} f_{a}\frac{
\partial f'_{b}}{\partial v'^{\nu}}
\right)\,d{\bf v}',
\end{eqnarray}
\begin{eqnarray}
\label{hb12}
K^{\mu \nu} = 2 \pi G^2 \frac{1}{u} \ln\Lambda \left(\delta^{
\mu \nu} - \frac{u^{\mu}u^{\nu}}{u^2}\right),
\end{eqnarray}
where $\textbf{u} = \textbf{v} - \textbf{v}'$ is the relative velocity
of the particles involved in an encounter and $\ln \Lambda = \int_0^{+
\infty}\,dk/k$ is the gravitational Coulomb factor.  We have set
$f_{a}'=f_{a}({\bf r},{\bf v}',t)$ assuming that the encounters can be
treated as local. The Landau-Poisson system monotonically increases
the entropy while conserving the mass $M_{a}$ of each species and the
total energy $E$. The linearly dynamically stable stationary state is
the mean-field Maxwell-Boltzmann distribution (\ref{hb9}) that is a
local maximum of Boltzmann entropy at fixed mass and energy
(microcanonical description). Therefore, the statistical equilibrium
states of Hamiltonian and Brownian systems have the same distribution
profiles but their stability may differ in case of ensemble
inequivalence. This occurs in particular when the caloric curve
presents turning points as discussed in Sec. \ref{sec_thermo}.

\section{Virial theorem in $d$ dimensions}
\label{sec_virial}

\subsection{The general expression}

Let us establish the Virial theorem associated with the stochastic
equations (\ref{hb1}). For convenience, we shall assume that the
friction coefficient $\xi$ is the same for all the particles but
this assumption can  be relaxed easily. The moment of inertia tensor
is defined by
\begin{equation}
I_{ij}=\sum_{\alpha}m_{\alpha}x_{i}^{\alpha}x_{j}^{\alpha}.
\label{exV1}
\end{equation}
We introduce the kinetic energy tensor
\begin{equation}
K_{ij}={1\over 2}\sum_{\alpha}m_{\alpha}{\dot x}_{i}^{\alpha}{\dot x}_{j}^{\alpha},
\label{exV2}
\end{equation}
and the potential energy tensor
\begin{eqnarray}
W_{ij}=G\sum_{\alpha\neq\beta} m_{\alpha}m_{\beta}{x_{i}^{\alpha}
(x_{j}^{\beta}-x_{j}^{\alpha})\over |{\bf r}_{\beta}
-{\bf r}_{\alpha}|^{d}}
=-{1\over 2}G \sum_{\alpha\neq\beta}
m_{\alpha}m_{\beta}{(x_{i}^{\alpha}-x_{i}^{\beta})
(x_{j}^{\alpha}-x_{j}^{\beta})\over
|{\bf r}_{\beta}-{\bf r}_{\alpha}|^{d}}, \label{exV3}
\end{eqnarray}
where the second equality results from simple algebraic manipulations
obtained by exchanging the dummy variables $\alpha$ and $\beta$.
Taking the second derivative of Eq.~(\ref{exV1}), using the equation
of motion (\ref{hb1}), and averaging over the noise and on
statistical realizations, we get
\begin{eqnarray}
{1\over 2} \langle {\ddot I}_{ij}\rangle +{1\over 2}\xi \langle {\dot
I}_{ij}\rangle=2\langle K_{ij}\rangle+\langle W_{ij}\rangle -{1\over 2}\oint
(P_{ik}x_{j}+P_{jk}x_{i})\,dS_{k}. \label{exV4}
\end{eqnarray}
For the sake of generality, we have accounted for the presence of a
box enclosing the system and included the Virial of the pressure force
$F_{i}=P_{ik}\Delta S_{k}$ on the boundary of the box: $(\ddot
I_{ij})_{box}=\sum_{box}\sum_{\alpha}(F_{i}^{\alpha}x_{j}^{\alpha}
+F_{j}^{\alpha}x_{i}^{\alpha})=\sum_{box}\sum_{\alpha}(P_{ik}^{\alpha}
x_{j}^{\alpha}+P_{jk}^{\alpha}x_{i}^{\alpha})\,\Delta S_{k}$. The
scalar Virial theorem is obtained by contracting the indices yielding
\begin{eqnarray}
{1\over 2} \langle {\ddot I}\rangle +{1\over 2}\xi \langle {\dot I}\rangle =2\langle K\rangle +\langle W_{ii}\rangle -\oint P_{ik}x_{i}dS_{k},
\label{exV5}
\end{eqnarray}
where
\begin{eqnarray}
I=\sum_{\alpha}m_{\alpha}r_{\alpha}^{2},\qquad K={1\over 2}
\sum_{\alpha}m_{\alpha}v_{\alpha}^{2},
\label{exV6}
\end{eqnarray}
are the moment of inertia and the kinetic energy. On the other hand,
\begin{eqnarray}
W_{ii}=-{1\over 2}G\sum_{\alpha\neq\beta} {m_{\alpha}m_{\beta}\over
|{\bf r}_{\beta}-{\bf r}_{\alpha}|^{d-2}}.
\label{exV7}
\end{eqnarray}

\subsection{The trace of the potential energy tensor}

For $d\neq 2$, we find that
\begin{eqnarray}
W_{ii}=(d-2)W
\label{exV8}
\end{eqnarray}
where $W$ is the potential energy
\begin{eqnarray}
W=-{G\over 2(d-2)}\sum_{\alpha\neq\beta}{m_{\alpha}m_{\beta}\over
|{\bf r}_{\beta}-{\bf r}_{\alpha}|^{d-2}}. \label{exV9}
\end{eqnarray}
In that case, the scalar Virial theorem reads
\begin{eqnarray}
{1\over 2} \langle {\ddot I}\rangle +{1\over 2}\xi \langle {\dot I}\rangle =2\langle K\rangle +(d-2)\langle W\rangle -\oint P_{ik}x_{i}dS_{k}.
\label{exV5bis}
\end{eqnarray}

For $d=2$, we have the simple result
\begin{eqnarray}
W_{ii}=-{1\over 2}G\sum_{\alpha\neq\beta}m_{\alpha}m_{\beta}.
\label{exV10}
\end{eqnarray}
For equal mass particles,
\begin{eqnarray}
W_{ii}=-{1\over 2}G N(N-1)m^{2}, \label{exV11}
\end{eqnarray}
which reduces to
\begin{eqnarray}
W_{ii}\simeq -{1\over 2}G M^{2},
\label{exV11b}
\end{eqnarray}
for $N\gg 1$. We also note that
$\sum_{\alpha\neq\beta}m_{\alpha}m_{\beta}=
M^{2}-\sum_{\alpha}m_{\alpha}^{2}$. For typical mass distributions,
the first term is of order $\sim N^{2} m^{2}$ and the second of
order $N m^{2}$ (where $m$ is a typical mass). Therefore, in the
thermodynamic limit $N\rightarrow +\infty$, we get Eq.
(\ref{exV11b}) even for a multicomponent system \cite{sopik}.

\subsection{The equilibrium state}

At equilibrium, the Virial theorem (\ref{exV5}) reduces to
\begin{eqnarray}
2\langle K\rangle +\langle W_{ii}\rangle =\oint P_{ik}x_{i}dS_{k}.
\label{exV12}
\end{eqnarray}
For an unbounded domain ($P=0$), we get $2\langle K\rangle +\langle W_{ii}\rangle =0$ (more precisely, $2\langle K\rangle=-(d-2)\langle W\rangle$ for $d\neq 2$ and  $\langle K\rangle=GM^{2}/4$ for $d=2$ in the approximation (\ref{exV11b})). If the system is at statistical equilibrium, then $\langle K\rangle={d\over 2}Nk_{B}T$ and $P_{ij}=p\delta_{ij}$ with $p=\sum_{s}\rho_{s}k_{B}T/m_{s}$ (where $s$ labels the species of particles) \cite{sopik}. If $p_{b}$ is constant  on the boundary of the box, the pressure term becomes
\begin{eqnarray}
\oint p_{b} {\bf r}\cdot d{\bf S}=p_{b}\oint {\bf r}\cdot d{\bf S}=p_{b}\int \nabla\cdot {\bf r} \,dV=d p_{b} V.
\label{exV13}
\end{eqnarray}
More generally, we define $P\equiv {1\over dV}\oint p {\bf r}\cdot
d{\bf S}$. In that case, the equilibrium Virial theorem becomes
\begin{eqnarray}
d N k_{B} T+\langle W_{ii}\rangle =d PV.
\label{exV14}
\end{eqnarray}
For an ideal gas without self-gravity ($W=0$), we recover the perfect gas law
\begin{eqnarray}
PV=N k_{B} T.
\label{exV15}
\end{eqnarray}
Alternatively, for a self-gravitating gas in two dimensions, using Eq. (\ref{exV10}), we get the exact equation of state
\begin{equation}
PV=Nk_{B}(T-T_{c}),
\label{exV17n}
\end{equation}
with the exact critical temperature
\begin{equation}
k_{B}T_{c}={G \sum_{\alpha\neq\beta}m_{\alpha}m_{\beta}\over 4N}.
\label{exV18}
\end{equation}
For equal mass particles, we get
\begin{equation}
k_{B}T_{c}={Gm^{2}\over 4}(N-1),
\label{exV18b}
\end{equation}
and in the $N\rightarrow +\infty$ limit
\begin{equation}
k_{B}T_{c}={GM^{2}\over 4N}.
\label{exV18bb}
\end{equation}
This relation is also valid for a multicomponent system in the
thermodynamic limit (see remark after Eq. (\ref{exV11b}))
\cite{sopik}. The equation of state (\ref{exV17n}) has been obtained
by Lynden-Bell \& Katz \cite{lbkatz}, Padmanabhan \cite{paddy} and
Chavanis \& Sire \cite{lang} using different methods.

\subsection{The critical dimension $d=4$ for Hamiltonian systems}

For Hamiltonian systems ($D=\xi=0$), the total energy $E=K+W$ is
conserved. Thus, in an unbounded domain ($P=0$), the Virial theorem
(\ref{exV5}) can be written for $d\neq 2$:
\begin{eqnarray}
{1\over 2} {\ddot I}=2 K +(d-2) W=(4-d)K+(d-2)E=2E+(d-4)W.
\label{exV19}
\end{eqnarray}
This is the extension of the usual Virial theorem in $d$ dimensions
(this equation is exact without averages). We note that the
dimension $d=4$ is {\it critical} as was also noticed in
\cite{fermionsd} using different arguments. In that case, ${\ddot
I}=4E$ which yields after integration $I=2Et^{2}+C_{1}t+C_{2}$. For
$E>0$, $I\rightarrow +\infty$ indicating that the system evaporates
(there can be equilibrium states if the system is confined within a
box). For $E<0$, $I$ goes to zero in a finite time, indicating that
the system forms a Dirac peak in a finite time. More generally, for
$d\ge 4$, since $(d-4)W\le 0$, we have $I\le 2Et^{2}+C_{1}t+C_{2}$
so that the system forms a Dirac peak in a finite time if $E<0$
(this result remains valid for box-confined systems). Therefore,
self-gravitating systems with $E<0$ are not stable in a space of
dimension $d\ge 4$. The study in \cite{fermionsd} indicates that
this observation remains true even if quantum effects (Pauli
exclusion principle for fermions) are taken into account. This is a
striking result because quantum mechanics stabilizes matter against
gravitational collapse in $d<4$ \cite{fermions}. For $2<d\le 4$,
since $(d-4)W\ge 0$ we conclude, according to Eq. (\ref{exV19}),
that the system evaporates if $E>0$ (there can be equilibrium states
if the system is confined within a box). An equilibrium is possible,
but not compulsory, if $E<0$ (the case of statistical equilibrium is
discussed in Sec. \ref{sec_thermo}). Finally, for $d<2$, since
$W>0$, the energy is necessarily positive ($E>0$). Since $(d-4)W<0$,
an equilibrium state is possible. In $d=2$, the Virial theorem
becomes with the approximation (\ref{exV11b}):
\begin{eqnarray}
{1\over 2} {\ddot I}=2 K -{GM^{2}\over 2}.
\label{exV20}
\end{eqnarray}
Since $K\ge 0$, an equilibrium state is possible.

\subsection{The critical dimension $d=2$ for Brownian systems}

We consider a self-gravitating Brownian gas in the strong friction
limit $\xi\rightarrow +\infty$. To leading order in $1/\xi$, the
$N$-body distribution is given by
\begin{equation}
P_{N}({\bf r}_{1},{\bf v}_{1},...,{\bf r}_{N},{\bf
v}_{N},t)=e^{-\beta \sum_{\alpha}m_{\alpha}{v_{\alpha}^{2}\over
2}}\Phi_{N}({\bf r}_{1},...,{\bf r}_{N},t)+O(\xi^{-1}),
\label{exV21}
\end{equation}
as can be deduced from Eq. (\ref{hb2}) by canceling the term in brackets.
From this expression, we find that $P_{ij}=p\delta_{ij}$ with
$p=\sum_{s}\rho_{s}k_{B}T/m_{s}$ and $\langle K_{ij}\rangle ={1\over
2}Nk_{B}T\delta_{ij}$ even for the out-of-equilibrium problem. From
Eq. (\ref{exV4}), we obtain the overdamped Virial theorem for a
self-gravitating Brownian gas
\begin{equation}
{1\over 2}\xi \langle {\dot I}_{ij}\rangle=\langle W_{ij}\rangle +Nk_{B}T\delta_{ij}-{1\over
2}\oint p(\delta_{ik}x_{j}+\delta_{jk}x_{i})\,dS_{k}.
\label{exV22}
\end{equation}
We can obtain this result in a different manner. In the strong
friction limit $\xi\rightarrow +\infty$, the inertial term in
Eq. (\ref{hb1}) can be neglected so that the stochastic equations of
motion reduce to
\begin{equation}
\dot x_{i}^{\alpha}=\mu_{\alpha}m_{\alpha}\sum_{\beta\neq \alpha}
{Gm_{\beta}(x_{i}^{\beta}-x_{i}^{\alpha})\over |{\bf
r}_{\beta}-{\bf
r}_{\alpha}|^{d}}+\sqrt{2D'_{\alpha}}R_{i}^{\alpha}(t),
\label{exV23}
\end{equation}
where $D'_{\alpha}=k_{B}T\mu_{\alpha}$ is the diffusion coefficient
in position space and $\mu_{\alpha}=(\xi m_{\alpha})^{-1}$ is the
mobility. Taking the derivative of the tensor of inertia
(\ref{exV1}) and using Eq.~(\ref{exV23}), we get
\begin{equation}
{\dot I}_{ij}={2\over\xi} W_{ij}+\sum_{\alpha}
m_{\alpha}\sqrt{2D'_{\alpha}}\lbrack
x_{i}^{\alpha}R_{j}^{\alpha}(t)+x_{j}^{\alpha}R_{i}^{\alpha}(t)\rbrack.
\label{exV24}
\end{equation}
Averaging over the noise using $\langle
x_{i}^{\alpha}R_{j}^{\alpha}\rangle=\sqrt{2D'_{\alpha}}\delta_{ij}$
and on statistical realizations, we recover Eq. (\ref{exV22}).  The scalar
Virial theorem obtained by contracting the indices reads
\begin{equation}
{1\over 2}\xi \langle {\dot I}\rangle =dNk_{B}T+\langle W_{ii}\rangle -dPV.
\label{exV25}
\end{equation}
In particular, in $d=2$, using Eq.~(\ref{exV10}) we find that
\begin{equation}
{1\over 2}\xi \langle {\dot I}\rangle =2Nk_{B}(T-T_{c})-2PV,
\label{exV26}
\end{equation}
where $T_{c}$ is defined in Eq. (\ref{exV18}). For $T>T_{c}$, an
equilibrium state is possible in a box, while the system evaporates
($I\rightarrow +\infty$) in an unbounded domain ($P=0$). For
$T<T_{c}$, the system creates a Dirac peak in a finite time ($I=0$ at
$t=t_{end}$). These dynamical evolutions are studied precisely in
\cite{sc,virial} by solving the Smoluchowski-Poisson system.

\subsection{The generalized Einstein relation including self-gravity in $d=2$}

In an unbounded
domain $(P=0)$, the Virial relation (\ref{exV26}) reduces to
\begin{equation}
{1\over 2}\xi \langle {\dot I}\rangle =2Nk_{B}(T-T_{c}).
\label{exV27}
\end{equation}
Defining the spatial dispersion of the particles by
\begin{equation}
\langle r^{2}\rangle={\langle \sum m_{\alpha}r_{\alpha}^{2}\rangle \over \sum m_{\alpha}}={\langle I\rangle\over M},
\label{exV28}
\end{equation}
we can rewrite the foregoing relation in the form
\begin{equation}
{d\langle r^{2}\rangle\over dt}={4Nk_{B}T\over\xi M}(1-T_{c}/T).
\label{exV29}
\end{equation}
Integrating this relation, we obtain
\begin{equation}
\langle r^{2}\rangle (t)={4Nk_{B}T\over\xi M}(1-T_{c}/T)t+\langle r^{2}\rangle_{0}.
\label{exV30}
\end{equation}
This relation suggests to introducing an effective diffusion coefficient
\begin{equation}
D_{eff}(T)={Nk_{B}T\over\xi M}(1-T_{c}/T),
\label{exV31}
\end{equation}
so that
\begin{equation}
\langle r^{2}\rangle (t)=4D_{eff}(T)t+\langle r^{2}\rangle_{0}.
\label{exV32}
\end{equation}
For $T\gg T_{c}$ when gravitational effects become negligible with
respect to thermal motion, the diffusion coefficient is given by the
celebrated Einstein formula $D=N k_{B}T/\xi M$. However,
Eq.~(\ref{exV31}) shows that the diffusion is less and less
effective as temperature decreases and gravitational effects come
into play. {\it Therefore, relation (\ref{exV31}) provides a
generalization of the Einstein relation to the case of
self-gravitating Brownian particles in $d=2$.} For $T>T_{c}$, we
have a diffusive motion (evaporation) with an effective diffusion
coefficient depending linearly on the distance $(T-T_{c})$ to the
critical temperature. For $T=T_{c}$, the effective diffusion
constant vanishes $D(T_{c})=0$ so that the moment of inertia is
conserved. Finally, for $T<T_{c}$ the effective diffusion
coefficient is negative, implying finite time collapse. In
particular, $\langle r^{2}\rangle=0$ for $t_{end}=\langle
r^{2}\rangle_{0}/4|D(T)|$ (where $\langle r^{2}\rangle_{0}$ is
calculated from the center of mass). This corresponds to the
formation of a Dirac peak $\rho({\bf r})=M\delta({\bf r})$ at ${\bf
r}={\bf 0}$ containing the whole mass \cite{virial}.

\section{Thermodynamics of self-gravitating classical and quantum particles}
\label{sec_thermo}

In this section we present a gallery of caloric curves for
self-gravitating classical and quantum particles (fermions) at
statistical equilibrium in different dimensions of space. From the
series of equilibria, we can directly infer the thermodynamical
stability limits of the system in microcanonical and canonical
ensembles by using the turning point argument of Poincar\'e. We also
directly see the critical parameters (energy and temperature) below
which the system collapses.  The microcanonical ensemble (MCE) is
appropriate to isolated Hamiltonian systems (in which case the control
parameter is the energy $E$) and the stable states are maxima of
entropy $S$ at fixed mass $M$ and energy $E$. The canonical ensemble
(CE) is appropriate to dissipative systems like self-gravitating
Brownian particles or self-gravitating gaseous systems in contact with
a thermal bath of non-gravitational origin (in which case the control
parameter is the temperature $T$) and the stable states are minima of
free energy $F=E-TS$ at fixed mass $M$.  The series of equilibria
$\beta(E)$ (representing critical points of the thermodynamical
potentials) are the {\it same} for the two systems but their stability
(related to the sign of the second order variations of the
thermodynamical potentials) can be different.  A turning point of
energy is associated with a loss of microcanonical stability (for
Hamiltonian systems) and a turning point of temperature is associated
with a loss of canonical stability (for Brownian systems). Stability
can be regained at the next turning point if the curve turns in the
reversed direction (as is the case for fermions).  We refer to Katz
\cite{katz}, Padmanabhan \cite{paddy} and Chavanis
\cite{aa1,aa3,fermions} for more precise statements of these results
and illustrations. In the following, the caloric curves are given
for box-confined systems. This is necessary to have existence of
statistical equilibrium states in $d>2$ (otherwise the system
evaporates). We also briefly discuss the case of unbounded systems
in $d=1$ and $d=2$.

\subsection{Classical particles}
\label{gmffp}

\begin{figure}
\vskip0.5cm \centerline{
\psfig{figure=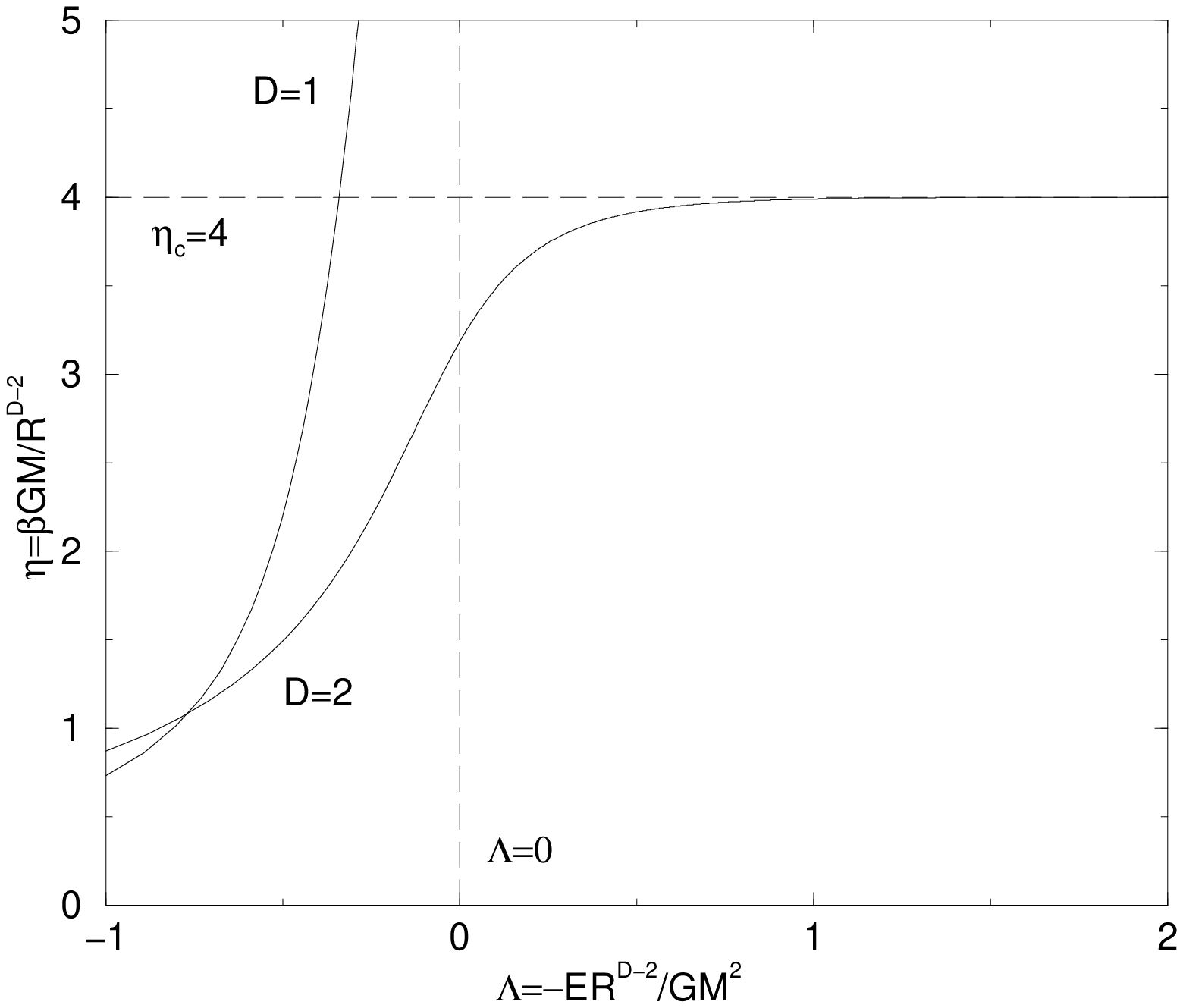,angle=0,height=6cm} \hspace{6pt}
\psfig{figure=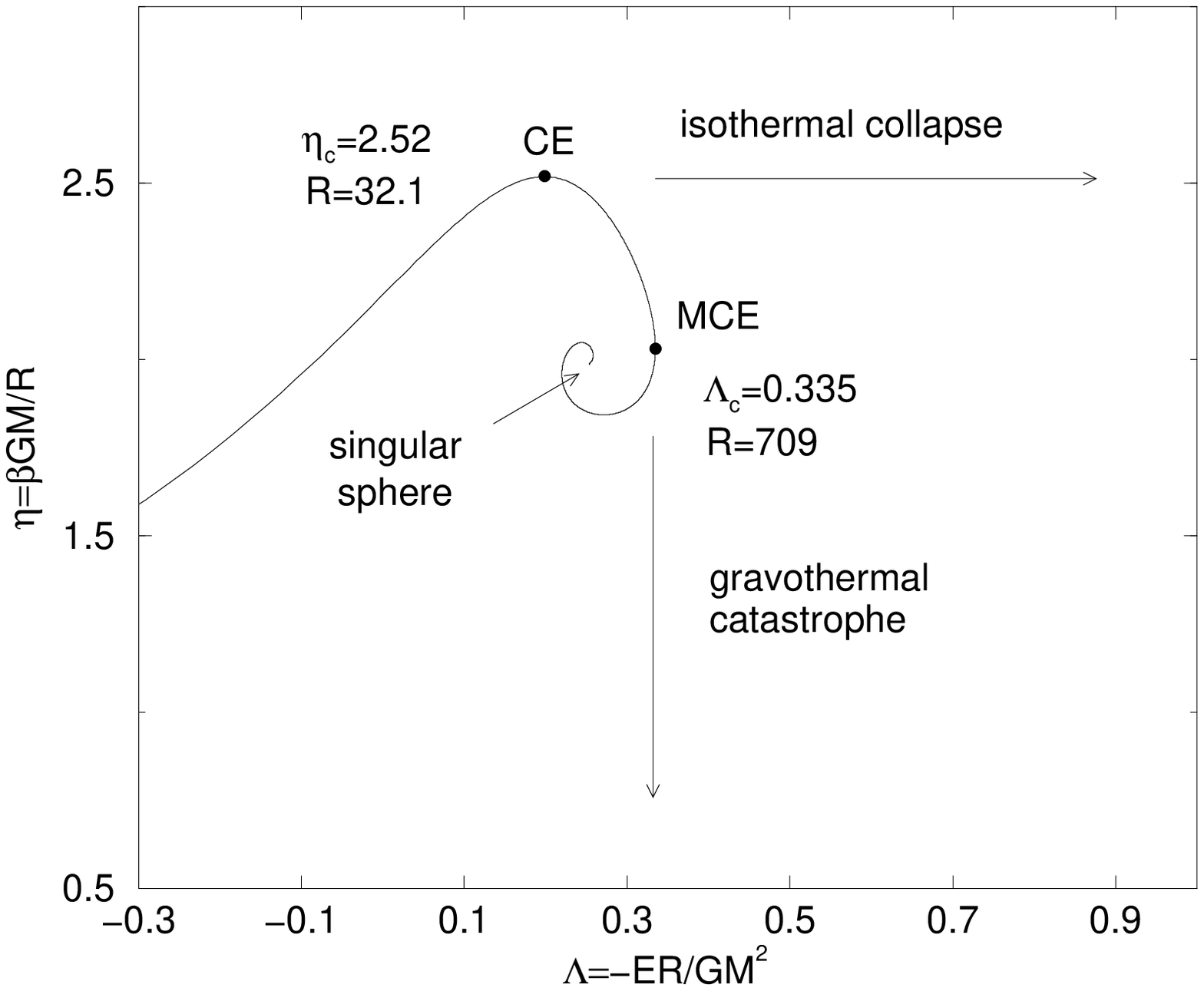,angle=0,height=6cm}} \caption{Series
of equilibria for classical isothermal spheres in $d=1$, $d=2$ and
$d=3$ dimensions. } \label{se12}
\end{figure}

In $d=1$, there exists statistical equilibrium states for any
accessible value of energy $E\ge 0$ in MCE and any value of
temperature $T$ in CE (see Fig. \ref{se12}-a). Since the series of
equilibria is monotonic, they are stable (global maxima of $S$ in
MCE and global minima of $F$ in CE) and the ensembles are
equivalent. There also exists stable statistical equilibrium states
in an unbounded domain. The density profile is known analytically
\cite{camm,sc,virial}. According to the Virial theorem, we have
$2K=W$ and $K={1\over 2}Nk_{B}T$ so that the caloric curve is simply
$E={3\over 2}Nk_{B}T$ \cite{virial}.

In $d=2$, there are statistical equilibrium states for any value of
energy $E$ in MCE and for $T>T_{c}=GMm/4k_{B}$ in CE (see Fig.
\ref{se12}-a). Their density profile is known analytically (see, e.g.,
\cite{sc}). Since the curve $\beta(E)$ is monotonic, they are stable
(global maxima of $S$ in MCE and global minima of $F$ in CE) and the
ensembles are equivalent. For $T\le T_{c}$ the self-gravitating
Brownian gas (canonical description) undergoes an isothermal
collapse and ultimately forms a Dirac peak \cite{sc}. In an
unbounded domain, there are stable statistical equilibrium states in
the microcanonical ensemble (Hamiltonian systems) for any value of
the energy \cite{perez}. According to the Virial theorem, we have
$K=GM^{2}/4$ and $K=Nk_{B}T$ so that these equilibrium states  have the
same temperature $T_{c}=GMm/4k_{B}$, independently on their energy
$E$. By contrast, in the canonical ensemble (Brownian systems),
there is no equilibrium state in an unbounded domain as discussed in
\cite{virial} and in Sec. 3.6.

\begin{figure}
\centerline{ \psfig{figure=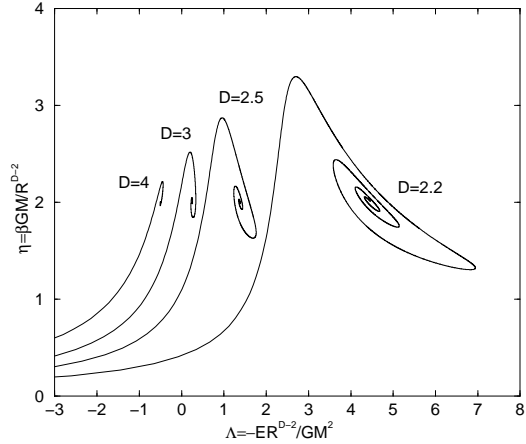,angle=0,height=6cm}}
\caption{Series of equilibria for classical isothermal spheres with
$2<d<10$.} \label{se3}
\end{figure}

In $d=3$, the series of equilibria forms a spiral (see Fig.
\ref{se12}-b). Since the series of equilibria presents turning points,
the ensembles are not equivalent. There exists stable statistical
equilibrium states in MCE for $E>E_{c}=-0.335GM^{2}/R$ (Antonov
energy); they have a density contrast ${R}\le 709$. There exists
stable statistical equilibrium states in CE for
$T>T_{c}=GMm/2.52k_{B}R$ (Emden temperature); they have a density
contrast ${R}\le 32.1$. They corresponds to long-lived {\it
metastable} states (local maxima of $S$ in MCE and local minima of $F$
in CE) whose lifetime $\sim e^{N}$ increases exponentially with the
number of particles \cite{metastable}. 
The states with density
contrast $32.1\le R\le 709$ are stable in MCE and unstable in CE; they
have negative specific heats. 
For $E<E_{c}$ a self-gravitating
Hamiltonian system (e.g. a globular cluster) undergoes a gravothermal
catastrophe \cite{lbwood}. There is first a self-similar collapse
leading to a finite time singularity (the central density becomes
infinite in a finite time) and ultimately the system forms a binary
star surrounded by a hot halo \cite{bt}. This structure has an
infinite entropy at fixed energy (see Appendix A of \cite{sc}).  For
$T<T_{c}$ a self-gravitating Brownian system undergoes an isothermal
collapse \cite{crs}.  There is first a self-similar collapse leading
to a finite time singularity \cite{sc} and ultimately the system forms
a Dirac peak in the post-collapse regime
\cite{post}. This structure has an infinite free energy (see
Appendix B of \cite{sc}). There is no equilibrium state (no maximum of
entropy or minimum of free energy) in an unbounded domain: the system
evaporates or collapses \cite{sc,virial}.

In $d\ge 3$, the phenomenology is the same as in $d=3$ (see Fig.
\ref{se3}) but we note,
for curiosity, that the spiral disappears for $d\ge 10$ \cite{sc}.
On the other hand for $d\ge 4$ and $E<E_c$, the system forms a Dirac
peak in the MCE instead of a ``binary star + hot halo'' structure
\cite{sc} (see Appendix A of \cite{sc}).

\subsection{Self-gravitating fermions}
\label{fer}

We now present the series of equilibria for the self-gravitating Fermi
gas (described by the Fermi-Dirac entropy) in different dimensions of
space and for different values of the degeneracy parameter $\mu$. The
degeneracy parameter can be viewed as a normalized Planck constant
$\mu\sim \hbar^{-d}$, as an effective inverse small-scale cut-off
$\mu\sim 1/\epsilon$ or as the system size $\mu\sim R^{d(4-d)/2}$ (the
classical case is recovered for $\hbar\rightarrow 0$,
$\epsilon\rightarrow 0$, $R\rightarrow +\infty$ or $\mu\rightarrow
+\infty$). We refer to
\cite{fermions,fermionsd,rieutord} for details.

\begin{figure}
\vskip0.5cm \centerline{
\psfig{figure=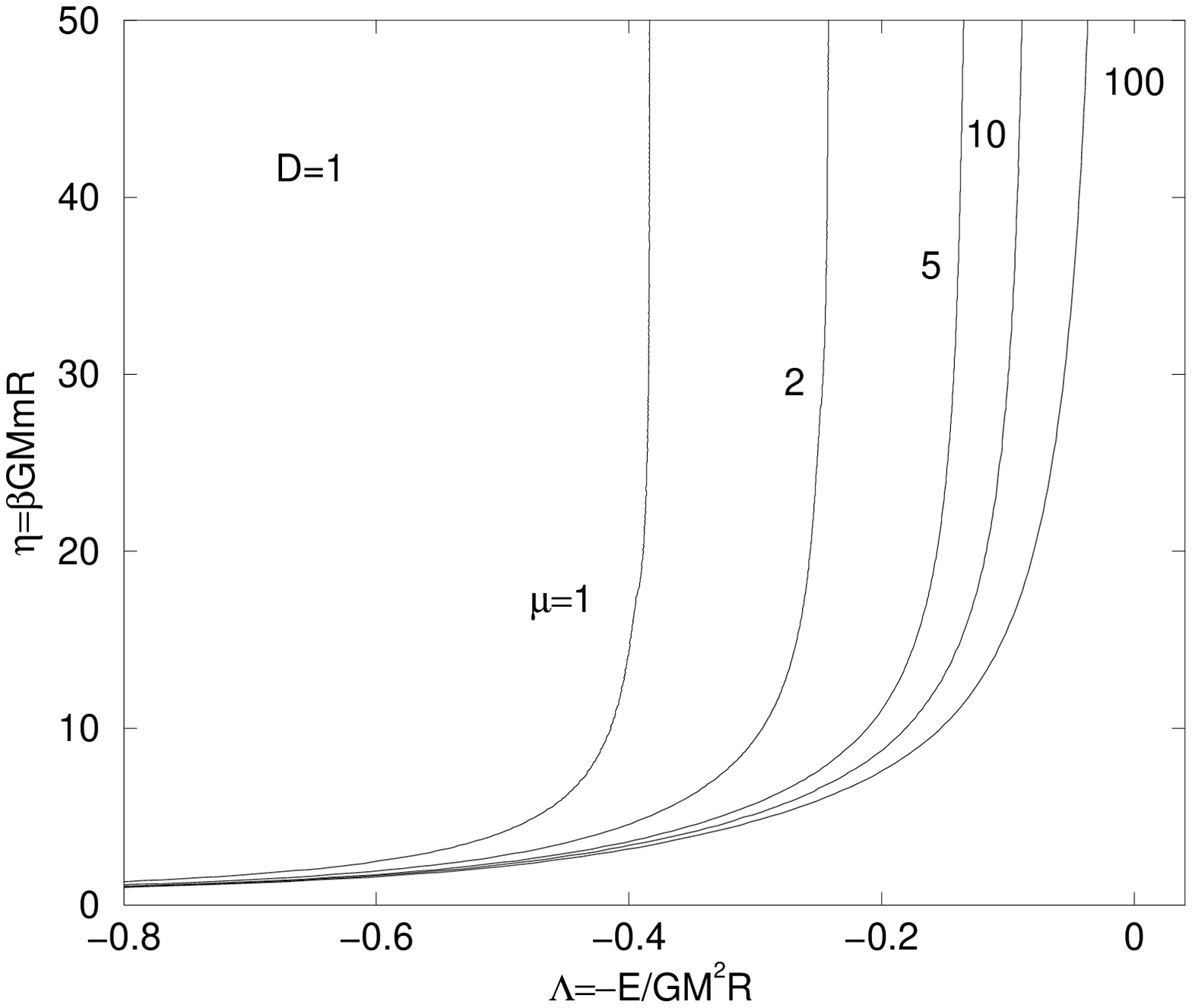,angle=0,height=6cm} \hspace{6pt}
\psfig{figure=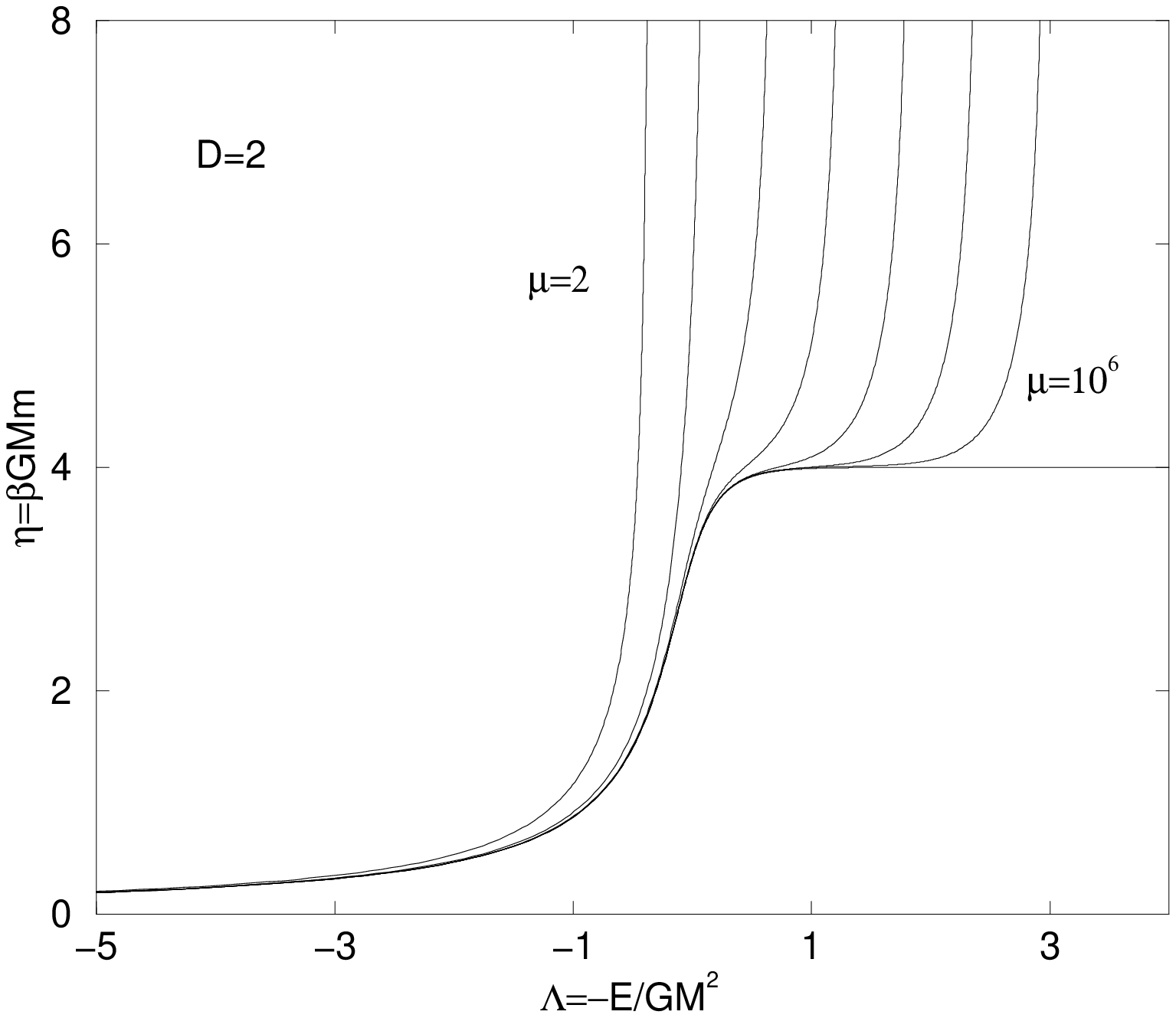,angle=0,height=6cm}} \caption{Series of
equilibria for an  isothermal gas of self-gravitating fermions  in
$d=1$ and $d=2$  dimensions for different values of the degeneracy
parameter (various system sizes). } \label{f12}
\end{figure}

In $d=1$, there exists statistical equilibrium states for any
accessible value of energy $E\ge E_{g}$ (where $E_{g}$ is
the ground state) in MCE and any value of
temperature $T$ in CE (see Fig. \ref{f12}-a). They are stable (global
maxima of $S$ in MCE and global minima of $F$ in CE).

In $d=2$, there exists statistical equilibrium states for any
accessible value of energy $E\ge E_{g}$  in MCE and any value of temperature $T$ in CE (see
Fig. \ref{f12}-b). They are stable (global maxima of $S$ in MCE and
global minima of $F$ in CE). For $\mu\rightarrow +\infty$,
$E_{g}\rightarrow -\infty$ and  we recover the classical
caloric curve displaying the critical temperature $T_{c}$. Below
$T_{c}$, the classical Brownian gas collapses and creates a Dirac
peak (``black hole'') \cite{sc}. When quantum mechanics is accounted
for, the ``black hole'' is replaced by a ``fermion ball'' (or white
dwarf star) consisting of a dense degenerate nucleus surrounded by a
dilute atmosphere \cite{fermionsd}.

\begin{figure}
\vskip0.5cm \centerline{
\psfig{figure=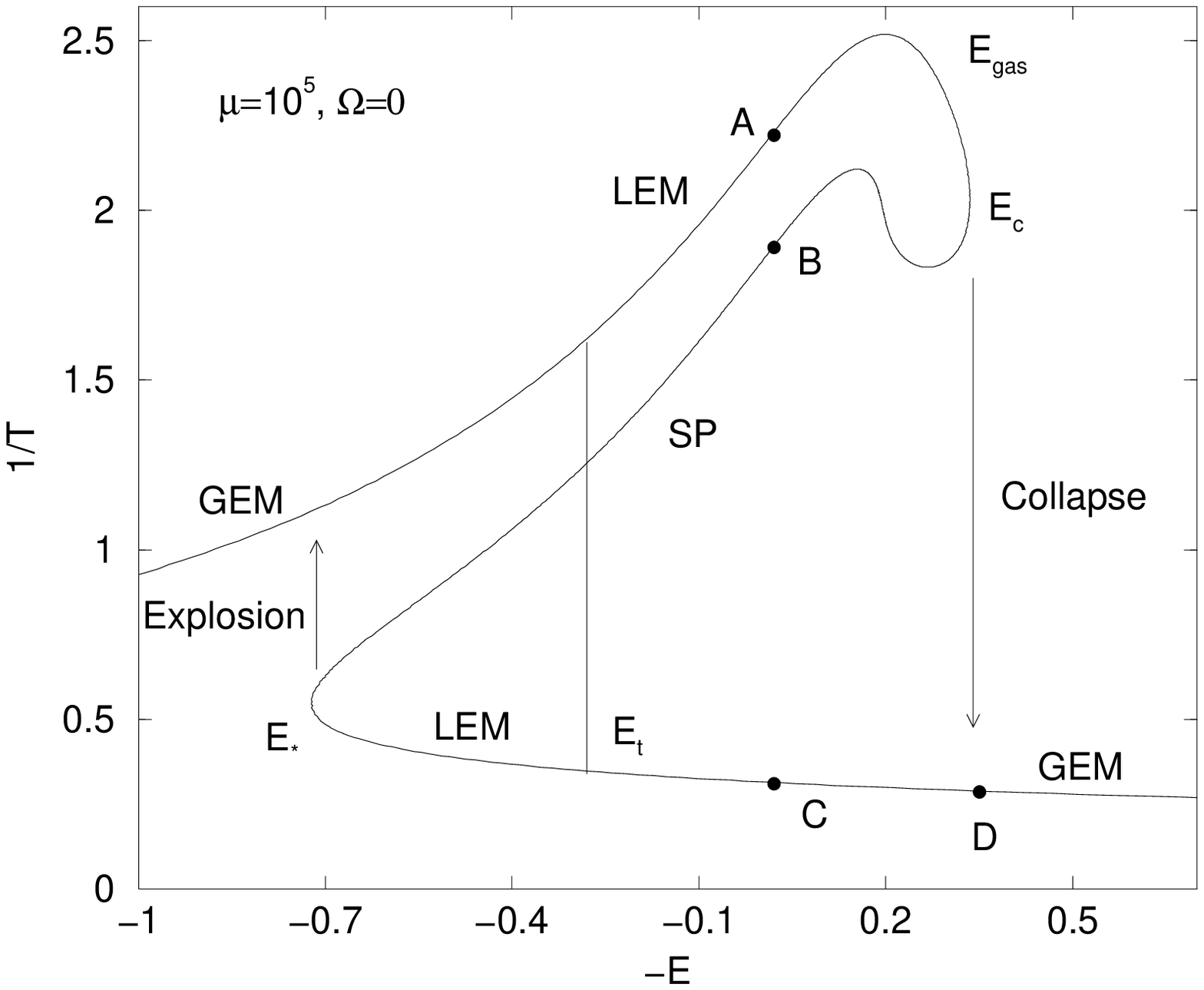,angle=0,height=6cm} \hspace{6pt}
\psfig{figure=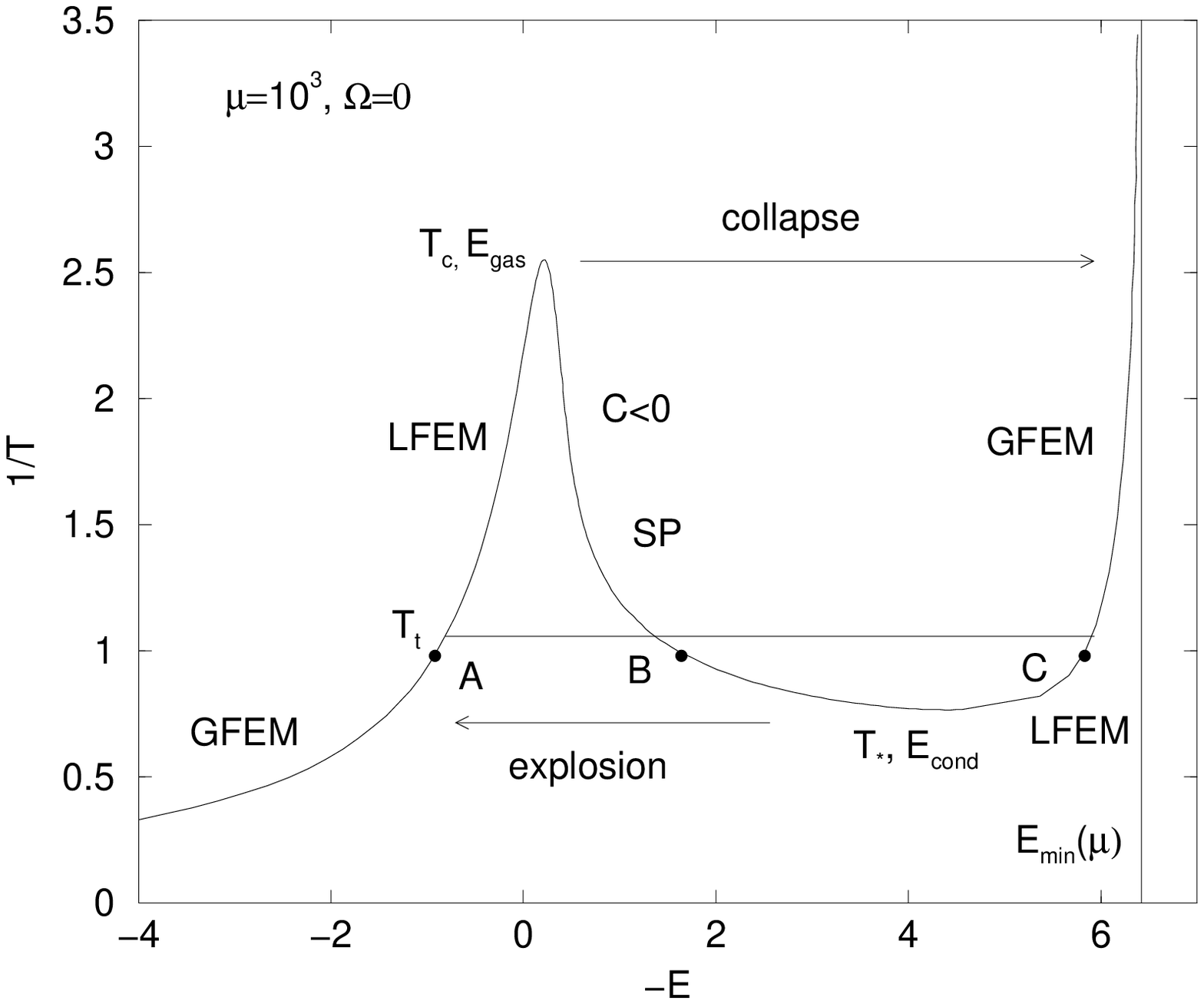,angle=0,height=6cm}} \caption{Series of
equilibria for an  isothermal gas of self-gravitating fermions  in
$d=3$ dimensions for large (left, $Z$-shape) and small (right,
$N$-shape) values of the degeneracy parameter. } \label{f3}
\end{figure}

In $d=3$, the nature of phase transitions depends on the value of the
degeneracy parameter \cite{fermions}. For high values of $\mu$ (low
small-scale cut-off), the series of equilibria has a $Z$-shape
structure resembling a dinosaur's neck (see Fig. \ref{f3}-a).
Starting from a gaseous configuration (upper branch) and decreasing
the energy, the system first passes from a stable gaseous phase
(global maximum of $S$) for $E>E_{t}$ to a metastable gaseous phase
(local maximum of $S$) for $E<E_{t}$. The microcanonical first order
phase transition at $E_{t}$ is avoided due to the long lifetime ($\sim
e^N$) of the metastable gaseous states. At the critical energy
$E_{c}$, the gaseous metastable phase disappears and the system
collapses (gravothermal catastrophe). However, the collapse is stopped
by quantum mechanics (Pauli exclusion principle) when the core of the
system becomes degenerate. Therefore, the system ends up in the
condensed phase (lower branch). If we now increase the energy the
system first passes from a stable condensed phase (global maximum of
$S$) for $E<E_{t}$ to a metastable condensed phase (local maximum of
$S$) for $E>E_{t}$. Again, the microcanonical first order phase
transition at $E_{t}$ is avoided due to the long lifetime of the
metastable condensed states. At the critical energy $E_{*}$, the
condensed metastable phase disappears and the system explodes and
returns to the gaseous phase. We can thus describe an hysteretic cycle
in the MCE. For low values of $\mu$ (large small-scale cut-off), the
series of equilibria $\beta(E)$ has a $N$-shape structure (see Fig.
\ref{f3}-b). Since there is no turning point of energy, 
all the states are stable in MCE (global maxima of $S$), for
Hamiltonian systems, and there is no special phase transition for this
value of degeneracy parameter, just a condensation (clustering) of the
system as $E$ decreases. Negative specific heats are possible in
MCE. By contrast, since the curve $E(\beta)$ has a $Z$-shape
structure, there is a phase transition in the CE, for Brownian
systems. Starting from a gaseous configuration (left branch) and
decreasing the temperature, the system first passes from a gaseous
stable phase (global minimum of $F$) for $T>T_{t}$ to a gaseous
metastable phase (local minimum of $F$) for $T<T_{t}$. The canonical
first order phase transition at $T_{t}$ is avoided due to the long
lifetime ($\sim e^N$) of the metastable gaseous states. At the
critical temperature $T_{c}$, the gaseous metastable phase disappears
and the system collapses (isothermal collapse). However, the collapse
is stopped by quantum mechanics (Pauli exclusion principle) when the
core of the system becomes degenerate.  Therefore, the system ends up
in the condensed phase (right branch). We get a ``fermion ball''
instead of a ``Dirac peak'' for classical particles. If we now
increase the temperature the system first passes from a stable
condensed phase (global minimum of $F$) for $T<T_{t}$ to a metastable
condensed phase (local minimum of $F$) for $T>T_{t}$. Again, the
canonical first order phase transition at $T_{t}$ is avoided due to
the long lifetime of the metastable condensed states. At the critical
temperature $T_{*}$, the condensed metastable phase disappears and the
system explodes and returns to the gaseous phase. We can thus describe
an hysteretic cycle in the CE
\cite{iso}. We note that when quantum effects are taken into account,
there exists an equilibrium state for any value of accessible energy
$E>E_{g}$ in MCE and any value of temperature in CE
\cite{fermions}. 

\begin{figure}[htbp]
\centerline{ \psfig{figure=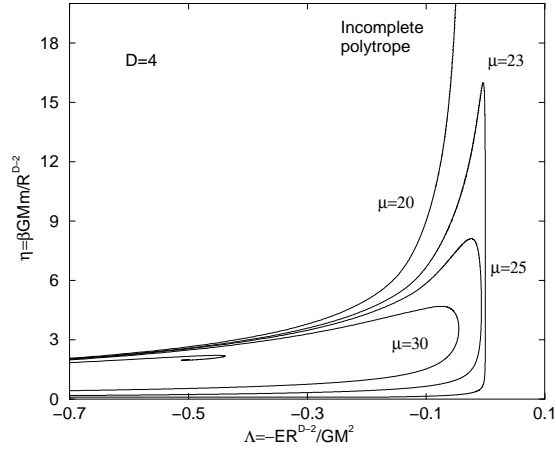,angle=0,height=6cm}}
\caption{Series of equilibria for an  isothermal gas of
self-gravitating fermions  in $d=4$  dimensions for different values
of the degeneracy parameter.  } \label{f4}
\end{figure}

In $d\ge 4$, the situation is different \cite{fermionsd}. In that
case, there is no equilibrium state for sufficiently small energies
and temperatures (see Fig. \ref{f4}). This is similar to the Antonov
instability for classical particles in $d=3$ but this now occurs for
fermions. Therefore, quantum mechanics cannot stabilize matter against
gravitational collapse in $d \ge 4$, contrary to what happens in
$d=3$. This is consistent with our previous observation that classical
white dwarf stars (the $T=0$ limit of the self-gravitating Fermi gas)
would be unstable in a space of dimension $d \ge 4$ \cite{lang}. This
is also similar to a result found by Ehrenfest \cite{ehrenfest} at the
molecular level (in Bohr's model). The case of relativistic white
dwarf stars in $d$ dimensions is treated in \cite{relD}.

\end{document}